\documentclass[10pt,journal,compsoc]{IEEEtran}
\usepackage{multirow}
\usepackage{tabularx}
\ifCLASSOPTIONcompsoc
  \usepackage[nocompress]{cite}
\else
  \usepackage{cite}
\fi
\ifCLASSINFOpdf
   \usepackage[pdftex]{graphicx}
   \graphicspath{{../pdf/}{../jpeg/}}
   \DeclareGraphicsExtensions{.pdf,.jpeg,.png}
\else
   \usepackage[dvips]{graphicx}
   \graphicspath{{../eps/}}
   \DeclareGraphicsExtensions{.eps}
\fi
\begin{document}
%
\title{An efficient sorting algorithm --- Ultimate Heapsort(UHS)}
%
%
%
%

\author{Feiyang Chen, Nan Chen, Hanyang Mao, Hanlin Hu}

%
%

\markboth{Chuangxinban Journal of computing, Mar~2018}%
{Shell \MakeLowercase{\textit{et al.}}: }
%



\IEEEtitleabstractindextext{%
\begin{abstract}
Motivated by the development of computer theory, sorting algorithm is emerging in an endless stream. Inspired by decrease and conquer method, we propose a brand new sorting algorithm—Ultimately Heapsort. The algorithm consists of two parts: building a heap and adjusting a heap. Through the asymptotic analysis and experimental analysis of the algorithm, the time complexity of our algorithm can reach $O(nlogn)$ under any condition.Moreover, its space complexity is only $O(1)$. It can be seen that our algorithm is superior to all previous algorithms.
\end{abstract}

}

\maketitle

\IEEEdisplaynontitleabstractindextext

%
\IEEEpeerreviewmaketitle

\IEEEraisesectionheading{\section{Introduction}\label{sec:introduction}}

%
%
%
%
\IEEEPARstart{S}{orting} algorithm is one of the most important research areas in computer science. It has a wide range of applications in computer graphics, computer-aided design, robotics, pattern recognition, and statistics. At present, quicksort is generally considered as the best choice in practical sorting applications. However, when we need to dynamically add and delete data during the sorting, we find that the quicksort can't exert its performance. Therefore, we design an efficient sorting algorithm---the ultimate heapsort(UHS), which offers us a better solution to this type of problem.

\section{Ultimate Heapsort(UHS)}
\subsection{Preliminary}
\par
We use heap data structure to implement our ultimate heapsort algorithm. The heap is a complete binary tree, and it is divided into a max-heap and a min-heap. The max-heap requires that the value of the parent node is greater than or equal to the value of the child node, and the min-heap is the opposite. According to the characteristic of the max-heap, we can know that the maximum value must be at the top of the heap, that is, the root node. With this, we can build an array into a max-heap. Here we take the max-heap as an example. The min-heap is similar. For our UHS algorithm, we use max-heaps.

\subsection{Main idea}
Here we will describe our UHS algorithm's main idea in detail. Firstly, we build the array as a max-heap, which is the initial heap. At this point we know that the top element of the heap is the maximum, that is, the first element of the array is the maximum value. Secondly, we swap the first element of the array with the last one, then the last one will be the maximum value, and then we update the heap with the last element removed, ensuring that the first one is at its maximum in the new heap. Finally, we repeat the above operation until only one element left.

\subsection{Design of the UHS}
\subsubsection{MAX-HEAPLFY}
In order to maintain the max-heap property, we call the procedure MAX-HEAPIFY. Its inputs are an array $A$ and an index $i$ into the array. When it is called, MAX- HEAPIFY assumes that the binary trees rooted at $LEFT(i)$ and $RIGHT(i)$ are max- heaps, but that A[i] might be smaller than its children, thus violating the max-heap property. MAX-HEAPIFY lets the value at $A[i]$ "float down" in the max-heap so that the subtree rooted at index $i$ obeys the max-heap property. 
	\begin{figure}[ht]

	\includegraphics[scale=0.32]{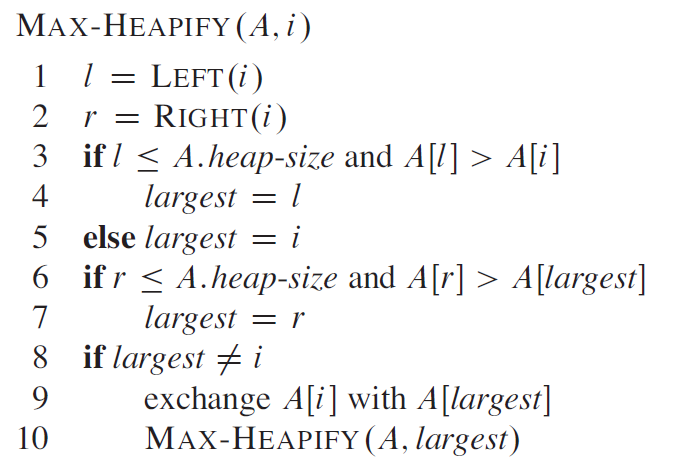}
	\label{fig:pathdemo}
\end{figure}
\par
At each step of MAX-HEAPLFY, the largest of the elements $A[i]$, $A[LEFT(i)]$, and $A[RIGHT(i)]$ is determined, and its index is stored in largest. If$ A[i] $is largest, then the subtree rooted at node $i$ is already a max-heap and the procedure terminates. Otherwise, one of the two children has the largest element, and $A[i]$ is swapped with $A[largest]$, which causes node $i$ and its children to satisfy the max-heap property. The node indexed by largest, however, now has the original value $A[i]$, and thus the subtree rooted at largest might violate the max-heap property. Consequently, we call MAX-HEAPIFY recursively on that subtree. 

\subsubsection{Building a heap}
We can use the procedure MAX-HEAPIFY in a bottom-up manner to convert an array $A[1..n]$, where $n = A.length$, into a max-heap. The elements in the subarray
 $A((\biggl\lfloor\frac{n}{2}\biggr\rfloor+1)..n)$ are all leaves of the tree, and so each is a $1$-element heap to begin with. The procedure BUILD-MAX-HEAP goes through the remaining nodes of the tree and runs MAX-HEAPIFY on each one. 
\begin{figure}[ht]

	\includegraphics[scale=0.32]{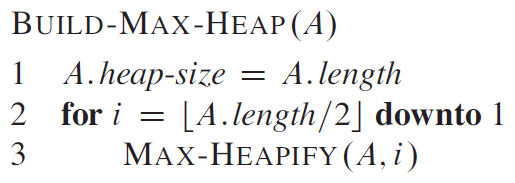}
	\label{fig:pathdemo}
\end{figure}
\subsubsection{Implementation}
The heapsort algorithm starts by using BUILD-MAX-HEAP to build a max-heap on the input array $A[1..n]$, where $n = A.length$. Since the maximum element of the array is stored at the root $A[1]$, we can put it into its correct final position by exchanging it with $A[n]$ . If we now discard node $n$ from the heap-and we can do so by simply decrementing $A.heap$-$size$-we observe that the children of the root remain max-heaps, but the new root element might violate the max-heap property. All we need to do to restore the max-heap property, however, is call MAX-HEAPIFY($A$,$1$), which leaves a max-heap in $A[1..n]$. The heapsort algorithm then repeats this process for the max-heap of size $n-1$ down to a heap of size $2$.
\begin{figure}[ht]

	\includegraphics[scale=0.32]{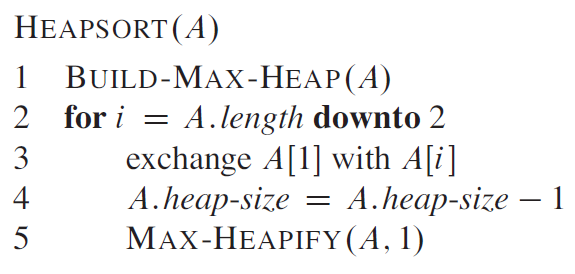}
	\label{fig:pathdemo}
\end{figure}

\section{Analysis}
\subsection{Running time analysis}
We have used mathematical methods to analyze the running time of UHS.\par
The maximum number of nodes in each layer during the build process is $n1 = ceil(n/(2^{h+1}))$, where $n$ and $h$ represent the number of nodes and the number of layers in the heap, respectively.
\begin{equation}
\sum_{h=0}^{\infty}=\frac{1}{x-1}
\end{equation}
\begin{equation}
\sum_{h=0}^{\infty}h*x^{h-1}*x^{2}=\frac{1}{(1-x)^{2}}*x^{2}
\end{equation}
\begin{equation}
\sum_{h=0}^{\infty}\frac{h}{2^{h+1}}=1
\end{equation}
\begin{equation}
\sum_{h=0}^{logn}n1*O(h)=O(n*\sum_{logn}^{h=0}\frac{h}{2^{h+1}})=O(n)
\end{equation}\par
Since the running time of resuming the heap is $O(logN)$ each time, a total of $N - 1$ times to restore the heap operation, plus the $N / 2$ times downward adjustment when the previous stack was built, the running time of each adjustment time is also $O(logN)$. The sum of the two operation times is also $O(N*logN)$. Therefore, the time complexity of heap sorting is $O(N*logN)$\par

\begin{table}[ht]
	\centering
	\caption{Running time analyze of 7 algorithms}
	\begin{tabular}{|l|l|l|}
		\hline	&Worst-case&Average-case/expected\\
		Algorithm&running time&running time\\
		\hline	Insertion sort&$\Theta(n^{2})$&$\Theta(n^{2})$\\
		\hline	Merge sort&$\Theta(nlogn)$&$\Theta(nlogn)$\\
		\hline	Quicksort&$\Theta(n^{2})$&$\Theta(nlogn)$ (expected)\\
		\hline Bucket sort&$\Theta(n^{2})$&$\Theta(n)$ (average-case)\\
		\hline	Radix sort&$\Theta(d(n+k))$&$\Theta(d(n+k))$\\
		\hline Bubble sort&$\Theta(n^{2})$&$\Theta(n^{2})$\\
		\hline	Heapsort&$\Theta(nlogn)$&$\Theta(nlogn)$\\
		
		\hline
	\end{tabular}
\end{table}
\par For the commonly used seven sorting algorithms, we have analyzed and compared the running time, including the worst-running time and expected running time . The comparison results are shown in the table 1.
\subsection{Space complexity analysis}
At the same time, we have also analyzed the space complexity of these seven sorting methods.
\begin{table}[ht]
	\centering
	\caption{Space complexity analysis of 7 algorithms}
	\begin{tabular}{|l|l|}	
		\hline Algorithm&Space complexty\\
		\hline	Insertion sort&$O(1)$\\
		\hline	Merge sort&$O(n)$\\
		\hline	Quicksort&$O(nlogn)$\\
		\hline Bucket sort&$O(n)$ \\
		\hline	Radix sort&$O(n+k)$\\
		\hline Bubble sort&$O(1)$\\
		\hline	Heapsort&$O(1)$\\
		
		\hline
	\end{tabular}
\end{table}

Table 2 shows the space complexity for each methods. Heap sort is in-sequence sorting algorithms, which means only a constant additional memory space is required in addition to the input array. Thus, UHS? performance was one of the best among all algorithms. The results suggested that UHS can increase the sorting speed without taking up more space.
\subsection{Stability analysis}
The structure of the heap is that the children of node $i$ are nodes $2*i$ and $2*i+1$ . The max heap requires that the parent node is greater than or equal to its $2$ child nodes, and the min heap requires that the parent node is less than or equal to its $2$ child nodes. In a sequence of length $n$, the process of heap sorting is to choose the largest (max heap) or smallest (min heap) values from the first $n/2$ and the total of $3$ values of its sub-nodes. The choice between these $3$ elements won?t affect the stability. However, when elements are selected for the parent nodes $n/2 - 1, n/2-2, ...1$, stability is destroyed. It is possible that the $n/2th$ parent node exchange swaps the next one, and the $n/2-1th$ parent does not swap the same subsequent element. Then the stability between these $2$ same elements is destroyed. Therefore, heap sorting is not a stable sorting algorithm.
\begin{table}[ht]
	\centering
	\caption{Stability analysis of 7 algorithms}
	\begin{tabular}{|c|c|}
		\hline Algorithm&Stability\\
		\hline	Insertion sort&YES\\
		\hline	Merge sort&YES\\
		\hline	Quicksort&NO\\
		\hline Bucket sort&YES \\
		\hline	Radix sort&YES\\
		\hline Bubble sort&YES\\
		\hline	Heapsort&NO\\
		
		\hline
	\end{tabular}
\end{table}
\subsection{Contrast}
According to the previous three parts, we can conclude that the advantage of heap sorting over other algorithms is that it can still maintain the running time of $O(nlogn)$ in the worst case. Compared to the quick-sorting algorithm that is efficient at the same time, it is an in-sequence sort and does not require an auxiliary array. And, because of the heap data structure, heap sorting is more convenient for dynamically adding and deleting from sorted arrays.

\section{Results}
We have proposed a new method for sorting algorithm, and named it ultimate heap sorting (UHS). Ultimate heap sorting is a comparison-based sorting algorithm. Its best, worst, and average running time are $O(n*logn)$. Heapsort is an in-place algorithm, but it is not a stable sort.
\par It divides its input into a sorted and an unsorted region, and it iteratively shrinks the unsorted region by extracting the largest element and moving that to the sorted region. The improvement consists of the use of a heap data structure rather than a linear-time search to find the maximum.
\par At the same time, according to experimental results, when the ultimate heap sorting algorithm is facing the situation that it needs to dynamically add data, its worst-case running time and average running time are improved compared with other sorting algorithms.


%

\ifCLASSOPTIONcaptionsoff
  \newpage
\fi

\end{document}